\pdfoutput=1
\documentclass[conference]{IEEEtran}
\IEEEoverridecommandlockouts
\usepackage{cite}
\usepackage{amsmath,amssymb,amsfonts}
\usepackage{algorithmic}
\usepackage{graphicx}
\usepackage{textcomp}
\usepackage{xcolor}
\usepackage{gensymb}
\def\BibTeX{{\rm B\kern-.05em{\sc i\kern-.025em b}\kern-.08em
 T\kern-.1667em\lower.7ex\hbox{E}\kern-.125emX}}
\begin{document}

\title{Robust High Mobility NLOS UE Beamforming Strategy for Gigantic MIMO

\thanks{This project has received funding from the European Union’s Horizon Europe research and innovation program under the Marie Skłodowska-Curie grant agreement No $101119643$, ultra-massive MIMO for future cell-free heterogeneous networks (MiFuture).}
}

\author{\IEEEauthorblockN{Josep R. Fernández Rull\textsuperscript{$\ast$,$\dagger$}, Liang Liu\textsuperscript{$\ast$}, Henrik Sjöland\textsuperscript{$\ast$,$\dagger$}, Juan Vidal Alegría\textsuperscript{$\ast$}}
\IEEEauthorblockA{\textsuperscript{$\ast$}Department of Electrical and Information Technology, Lund University, Lund, Sweden \\
\textsuperscript{$\dagger$}Ericsson AB, Lund, Sweden \\
\{josep\_ramon.fernandez\_rull, liang.liu, henrik.sjoland, juan.vidal\_alegria\}@eit.lth.se
} 
}
\maketitle
\begin{abstract}
Maintaining robust and stable communication links in high-mobility scenarios is challenging for time-division duplex (TDD) reciprocity-based gigantic MIMO systems due to the rapid channel variations, especially in non-line-of-sight (NLOS) conditions. This paper proposes a user equipment (UE) beamforming strategy that enables a robust link in high mobility without incurring additional pilot overhead. The proposed beamforming strategy consists of aligning the directional beamforming with the UE's travel axis. Our analysis shows the optimality of the proposed travel‑axis beamforming in minimizing the Doppler spread of the channel, which corresponds to an inverse metric for channel stability. This approach is further evaluated through simulations in a scattering-rich environment, representative of gigantic MIMO deployments. Numerical results confirm that movement‑aligned UE beamforming enhances link robustness, increases achievable data rates, and lowers pilot signaling requirements, thereby reducing UE power consumption. These findings highlight travel‑axis aligned UE beamforming as a promising technique for improving reliability in future high‑mobility wireless systems.

\end{abstract}


\section{Introduction}

%The increasing demand for higher data rates, together with the imperative to reduce energy consumption for sustainable development, has established the limitations of fifth-generation (5G) networks and motivated the move toward the sixth generation (6G) \cite{6g_and_beyond}. A key enabling technology of 5G is massive multiple‑input multiple‑output (massive MIMO), which provides large spatial degrees of freedom through beamforming and spatial multiplexing. To meet 6G requirements, more spectrum with larger bandwidth is needed, ranging from centimeter-wave (cmWave) to millimeter-wave (mmWave) and into terahertz (THz) bands \cite{millimeter_wave, 6g_spectrum}. The shorter wavelengths at these frequencies enable compact antenna arrays with many more elements \cite{UmMIMO_0}. In practice, massive MIMO in 5G is being extended with substantially higher antenna counts at base stations (BSs) and, increasingly, at user equipment (UE). For example, exploiting cmWave may require roughly $4$-$8$ times more BS antennas than sub‑$6$ GHz deployments to preserve the same physical aperture, assuming half-wavelength inter-element spacing \cite{FR3_4}, leading to gigantic MIMO \cite{gmimo}.

The increasing demand for higher data rates, together with the need for improved energy efficiency, has exposed the limitations of fifth-generation (5G) networks and motivated the move toward the sixth generation (6G) \cite{6g_and_beyond}. To meet 6G requirements, wider bandwidths across centimeter-wave (cmWave), millimeter-wave (mmWave), and sub-terahertz (THz) bands are required \cite{cmWave_THz}. The shorter wavelengths in these bands enable compact arrays with many more elements \cite{UmMIMO_0}. In practice, antenna configurations are being scaled up at both base stations (BSs) and user equipment (UE). For instance, exploiting cmWave may require roughly $4$-$8$ times more BS antennas than sub‑$6$GHz deployments to preserve the same physical aperture, assuming half-wavelength inter-element spacing \cite{FR3_4}, leading to gigantic MIMO \cite{gmimo}.

This continued scaling of massive MIMO brings user mobility to the forefront, as larger arrays and higher carrier frequencies lead to faster channel variations that undermine beam alignment and channel stability. In high‑mobility scenarios, the coherence time can become so short that conventional beamforming based on Channel State Information (CSI) becomes impractical without excessive computational complexity and signaling overhead \cite{HighMobility}. At the same time, higher carrier frequencies make it feasible to integrate compact antenna arrays at the UE, enabling UE‑side beamforming to improve the link budget and compensate for increased path loss \cite{beamforming}. However, the effectiveness of such directional transmission critically depends on maintaining beam alignment over time, which becomes increasingly challenging under mobility \cite{HighMobility}.

%At mmWave frequencies, UE‑side beamforming is facilitated by sparse propagation with only a few dominant paths, making directional steering effective, while still requiring continuous beam selection and tracking, often under a single‑RF‑chain constraint \cite{5g_beam_track,directional_beamforming}. In contrast, at cmWave frequencies, line‑of‑sight (LOS) components are generally less dominant, and signal propagation is characterized by richer scattering, requiring beamforming strategies that account for a larger number of contributing paths \cite{channel_cmWave,channel_cmWave2}. Under high‑mobility conditions, this richer multipath structure significantly increases beam‑management overhead, as beam updates must be performed more frequently to track the rapidly varying channel.

At mmWave frequencies, UE‑side beamforming is facilitated by sparse propagation with only a few dominant paths, making directional steering effective, while still requiring continuous beam selection and tracking \cite{5g_beam_track,directional_beamforming}. In contrast, cmWave channels exhibit weaker line‑of‑sight (LOS) components and richer scattering, requiring beamforming strategies that account for a larger number of contributing paths \cite{channel_cmWave,channel_cmWave2}. Under mobility, this richer multipath structure substantially increases the beam‑management burden, as more frequent updates are needed to track the rapidly varying channel.

Prior work on mmWave UE beamforming show that, under dominant LoS conditions, the UE beam can be designed using position and velocity information to form a broad, fixed receive beam that robustly covers the user’s predicted motion over a time interval, avoiding frequent beam updates \cite{mmWave_UE_beam}. Under NLOS mmWave scenarios, however, \cite{mmWave_NLOS} proposed an efficient codebook‑based beam training technique that addresses the presence of multiple comparable paths by formulating beam selection as a multi‑stage optimization problem with reduced training overhead. In cmWave, multiple studies document richer scattering and less dominant LOS \cite{channel_cmWave,channel_cmWave2}, which increases beam‑management overhead under mobility. Nevertheless, we are not aware of UE‑side beamforming schemes explicitly designed for high‑mobility cmWave systems.

To address the challenges previously mentioned, this study proposes a UE‑side beamforming technique in which the antenna beam is aligned with the axis of motion, thereby stabilizing the temporal evolution of the channel and reducing the need for frequent CSI updates. It is shown that such travel‑axis UE beamforming not only reduces signaling overhead, but also enhances overall communication performance by increasing data rates and providing more stable and reliable links in high‑mobility and NLOS conditions. Specifically, we provide a formal proof of the optimality of this approach in minimizing the Doppler spread of the propagation channel. Furthermore, a numerical comparison between this approach, omnidirectional transmission, and dominant eigenmode transmission is presented, demonstrating that maintaining a beam aligned with the UE trajectory offers clear advantages without the need to update beamforming weights given a movement distance.

The remainder of this paper is organized as follows: Section II presents the system model; the proposed solution is detailed in Section III; and Section IV provides numerical results and discussion. Finally, Section V concludes the paper and outlines potential directions for future lines of work. 

\vspace{2mm}

\section{System model}
A time-domain duplexing (TDD) system is considered in which an $N_\text{BS}$-antenna BS serves a moving $N_\text{UE}$-antenna UE through a narrowband channel. The exposition focuses on the uplink (UL) scenario, but the results are directly applicable to the downlink (DL) counterpart due to the angular reciprocity of the propagation channel. The received vector at the BS may be expressed as
\begin{equation}
\boldsymbol{y}=\boldsymbol{H}(t)\boldsymbol{w}_\text{UE} s + \boldsymbol{n},
\label{y}
\end{equation}
where $s$ is the transmitted symbol assuming single-layer transmission, $\boldsymbol{w}_{\text{UE}}$ is the UE beamforming vector, $\boldsymbol{n}\sim\mathcal{CN}(\boldsymbol{0},N_{0} \mathbf{I}_{N_{\text{BS}}})$ is the noise vector, and $\boldsymbol{H}(t)$ is the time-varying $N_{\text{BS}} \times N_{\text{UE}}$ channel matrix at time instant $t\geq 0$. Due to high-mobility, channel variations are considered within one coherence block, where channel estimation is performed at $t=0$. Moreover, the UE beamforming $\boldsymbol{w}_\text{UE}$ is pre-selected before channel estimation, and it is kept fixed throughout the whole coherence block.

NLOS conditions are assumed, such that the direct path is blocked, and only paths via scattering objects are present. In this case, the channel matrix may be expressed as
\begin{equation}
\boldsymbol{H}(t)= \boldsymbol{H}_{2} \text{diag} \left(\boldsymbol{\alpha}(t) \right) \boldsymbol{H}_{1}(t),
\label{eq:H_total}
\end{equation}
where $\boldsymbol{H}_{1}(t)$ is the time-varying channel matrix between the UE antennas and the scatterers, $\boldsymbol{H}_{2}$ is the fixed channel matrix between the scatterers and the BS antennas, and $\boldsymbol{\alpha}(t)$ is an $N_{\text{sc}}$-sized vector containing the complex baseband respond of each of the $N_{\text{sc}}$ scatterers.\footnote{The scattering response $\boldsymbol{\alpha}(t)$ is assumed time-dependent due to the dependency of the bistatic radar cross section (RCS) on the incidence angle.} To ensure validity of the results when the UE gets relatively close to the scatterers, we consider the non-uniform spherical wave (NUSW) near-field channel model, as presented in \cite{near_field}. Note that, as we go to higher frequencies, the likeliness of being in the near-field of the objects in the environment increases. The  channel between two points in space, $\boldsymbol{p}_{1}$ and $\boldsymbol{p}_{2}$, is thus given by
\begin{equation}
h(\boldsymbol{p}_1, \boldsymbol{p}_2)= \frac{1}{\sqrt{4 \pi \left|\left| \boldsymbol{p}_1-\boldsymbol{p}_2\right|\right|^2}}e^{- j \frac{2\pi}{\lambda}\left|\left| \boldsymbol{p}_1-\boldsymbol{p}_2\right|\right|},
\label{eq:NUSW}
\end{equation}
where $\lambda$ corresponds to the wavelength. Correspondingly, the entries of $\boldsymbol{H}_1(t)$ and $\boldsymbol{H}_2$, $\forall m \in \left\{1, \dots, N_{\text{sc}} \right\}$, $\forall n \in \left\{1, \dots, N_{\text{UE}} \right\}$ and $\forall \ell \in \left\{1, \dots, N_{\text{BS}} \right\}$, are given by
\begin{equation}
\begin{aligned}
[\boldsymbol{H}_1(t)]_{m,n} &= h(\boldsymbol{p}_m^\text{SC}, \boldsymbol{p}_n^\text{UE} (t))\\
[\boldsymbol{H}_2]_{\ell,m} &= h(\boldsymbol{p}_\ell^\text{BS}, \boldsymbol{p}_m^\text{SC})
,
\end{aligned}
\label{eq:H1H2}
\end{equation}
where $\boldsymbol{p}^\text{SC}_m$, $\boldsymbol{p}^\text{UE}_n(t)$, and $\boldsymbol{p}^\text{BS}_\ell$ are the position of the $m-$th scatterer, the $n-$th UE antenna element at time instant $t$, and the $l-$th BS antenna element, respectively. \\

\subsection{Beamforming considerations}
The BS applies a linear combining vector $\boldsymbol{w}_\text{BS}\in\mathbb{C}^{N_{\text{BS}}}$, leading to the post-processed symbol 
\begin{equation}
z=\boldsymbol{w}_\text{BS}^T\boldsymbol{y}\label{y}.
\end{equation}
Since the UE beamforming $\boldsymbol{w}_\text{UE}$ is applied at the UE prior to channel estimation, the BS can only observe and estimate the single-input multiple-output (SIMO) effective channel vector $\boldsymbol{h}_{\text{est}}=\boldsymbol{H}(0)\boldsymbol{ w}_\text{UE}$, estimated at $t=0$ through UL pilot signaling. For ease of exposition, perfect channel estimation is considered, but the gains of the proposed are not limited to that case. Assuming that the BS is unaware of the UE mobility, the best choice for its linear combining vector is given by maximum ratio combining (MRC). This choice maximizes the power of the received signal at $t=0$, and may be applied by selecting
\begin{equation}
\boldsymbol{w}_\text{BS} = \frac{\boldsymbol{h}_{\text{est}}^*}{\Vert\boldsymbol{h}_{\text{est}}\Vert}\triangleq\frac{\left(\boldsymbol{H}(0)\boldsymbol{ w}_\text{UE}\right)^*}{\left\Vert\boldsymbol{H}(0) \boldsymbol{w}_\text{UE} \right\Vert}.
\label{eq:w_rx}
\end{equation}
The instantaneous post-processing signal-to-noise ratio (SNR) at time $t$ is then given by
\begin{equation}
    \mathrm{SNR}\left(t\right)=\frac{E_s}{N_0}\left|\boldsymbol{w}_\text{BS}^T \boldsymbol{H}\left(t\right)\boldsymbol{w}_\text{UE} \right|^2,
\label{eq:SNR}
\end{equation}
where $E_s$ is the UE symbol power and $N_0$ is the noise power.

Both the UE and the BS are assumed to have their antennas organized into a uniform linear array (ULA) with half-wavelength spacing. For presentation clarity we focus on the 2-dimensional (2D) scenario, i.e., assuming that both ULAs are aligned in the 3rd dimension. The UE beamforming vector is then computed as
\begin{equation}
\boldsymbol{w}_\text{UE} = \frac{1}{\sqrt{N_\text{UE}}}\boldsymbol{a}_\text{UE}^*\left( \theta_\text{UE}\right),
\label{eq:w_UE}
\end{equation}
where $\theta_\text{UE}$ denotes the UE pointing direction measured from the array broadside, and $\boldsymbol{a}_\text{UE}\left( \theta_\text{UE}\right)$ is the traditional far-field steering vector given by
\begin{equation}
\boldsymbol{a}_\text{UE}\left( \theta_\text{UE}\right) = \left[1, e^{\text{j}\pi \sin\left(\theta_\text{UE}\right)}, \dots, e^{\text{j}\pi(N_{\text{UE}}-1) \sin\left(\theta_\text{UE}\right)} \right]^T.
\label{steering_vector}
\end{equation}
Back-lobe suppression is assumed at the UE array, meaning that the radiation toward the backward half-plane is set to zero. Under this convention, the UE can only steer its beam within a single $180^\circ$ sector, i.e.,
\begin{equation}
\theta_\text{UE}\in\left[ -\pi/2,\pi/2\right].
\label{eq:theta_ue}
\end{equation}
This forward-only behavior may be justified by the use of partly directive antenna elements, such as patch antennas, resulting in an effective directivity gain of approximately $3$dB.

%The main goal of this work is to characterize the most suitable pointing direction $\theta_{\text{UE}}$ that allows for increased channel coherence within the UE trajectory. Along this trajectory, the UE moves with a velocity vector that forms a known angle, $\theta_\text{mov}$, with respect to the array broadside. This angle is assumed to be fixed within the coherence block, and it can be obtained from on‑device information such as GPS, inertial sensors, or motion‑estimation algorithms.

\begin{figure}[htbp]
 \centering 
 \includegraphics[width=1\linewidth]{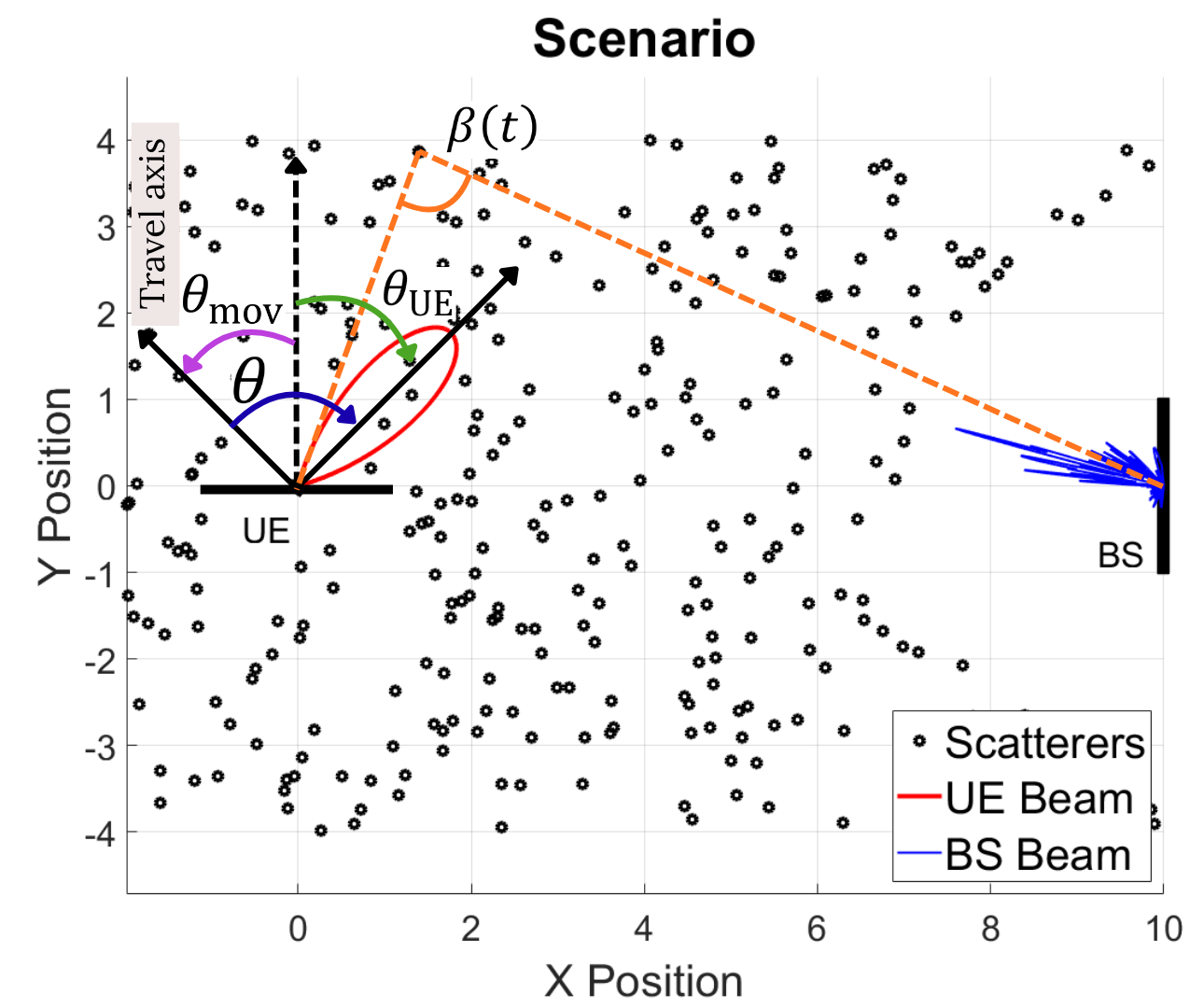}
 \caption{Illustration of the scenario.}
 \label{fig:scenario}
\end{figure}

The considered scenario is illustrated in Fig.~\ref{fig:scenario}, where $\theta_\text{mov}$ is the angle between the travel axis and the UE broadside. The angle $\theta_\text{mov}$ is assumed to be fixed within the coherence block, and known at the UE since it can be obtained from on‑device information such as GPS, inertial sensors, or motion‑estimation algorithms. The main goal of this work is to characterize the most suitable pointing direction $\theta_{\text{UE}}$ that allows for increased channel coherence within the UE trajectory.

\section{Travel-axis beamforming}
 
In this section, the main idea underlying the proposed solution is presented. The proposed approach relies on aligning the UE beamforming direction with its direction of motion, which is shown to be the most robust option in high-mobility scenarios. In this setting, the UE steers its transmit main lobe along the direction in which it is moving using the beamforming vector $\boldsymbol{w}_\text{UE}$ defined in \eqref{eq:w_UE}. 

%This choice of $\boldsymbol{w}_\text{UE}$ imposes a forward-looking radiation pattern, and backlobe suppression is assumed, meaning that all angles outside the forward half-plane are assigned zero gain. In practice, this results in an increase of approximately $3$dB in the forward gain due to the redistribution of the radiated power over a reduced angular region. 

%Under this model, the beamforming operation can be viewed as spatially weighting the angular spectrum of the transmitted energy with respect to the travel axis, such that multipath components whose angles lie close to the direction of motion are emphasized, while components associated with larger off‑axis angles are attenuated.

%In other words, this operation can be viewed as spatially weighting the angular spectrum of the transmitted energy with respect to the travel axis, such that multipath components whose angles lie close to the direction of motion are emphasized, while components associated with larger off‑axis angles are attenuated.

%This strategy is beneficial because UE motion converts angular dispersion into temporal channel variation. 
It is well known that the coherence time and the Doppler spread of the channel are inversely related \cite{molisch}. Thus, we next focus on demonstrating the optimality of travel-axis beamforming at the UE for increased channel coherence by solving an equivalent Doppler spread minimization problem. For a multipath component departing the UE at an angle $\theta$ relative to the UE travel-axis, the resulting Doppler shift is 
\begin{equation}
\nu\left(\theta\right)=\frac{\text{v}_\text{UE}}{c}f_c\cos\left( \theta\right),
\label{eq:doppler_1}
\end{equation}
where $\text{v}_\text{UE}$ denotes the UE speed, $c$ is the speed of light, and $f_c$ is the carrier frequency. This expression shows that the Doppler shift depends on the projection of the UE velocity onto the propagation direction of each transmitted path.

Let $\theta=\theta_{\text{UE}}-\theta_{\text{mov}}$ now denote the angle between the beam pointing direction and the travel-axis. We assume that the power radiated by the UE outside the main lobe is negligible. Moreover, the main lobe has beamwidth $B=2\gamma$, with $\gamma\leq \pi/2$, symmetric around the pointing direction, and we we consider a rich enough scattering environment such that all directions within the main lobe are associated with scattering objects. Two different cases are studied: one where the UE travel-axis is inside the main lobe, i.e., $\theta\in(-\gamma, \gamma)$, and the other where it is outside the lobe, i.e., $\theta\in[-\pi/2-\theta_{\text{mov}},-\gamma]\cup [\gamma, \pi/2-\theta_{\text{mov}}]$. Note that if the travel-direction points towards the backlobe area of the ULA we can equivalently consider the opposite direction.

\begin{figure}[htbp]
 \centering 
 \includegraphics[width=0.95\linewidth]{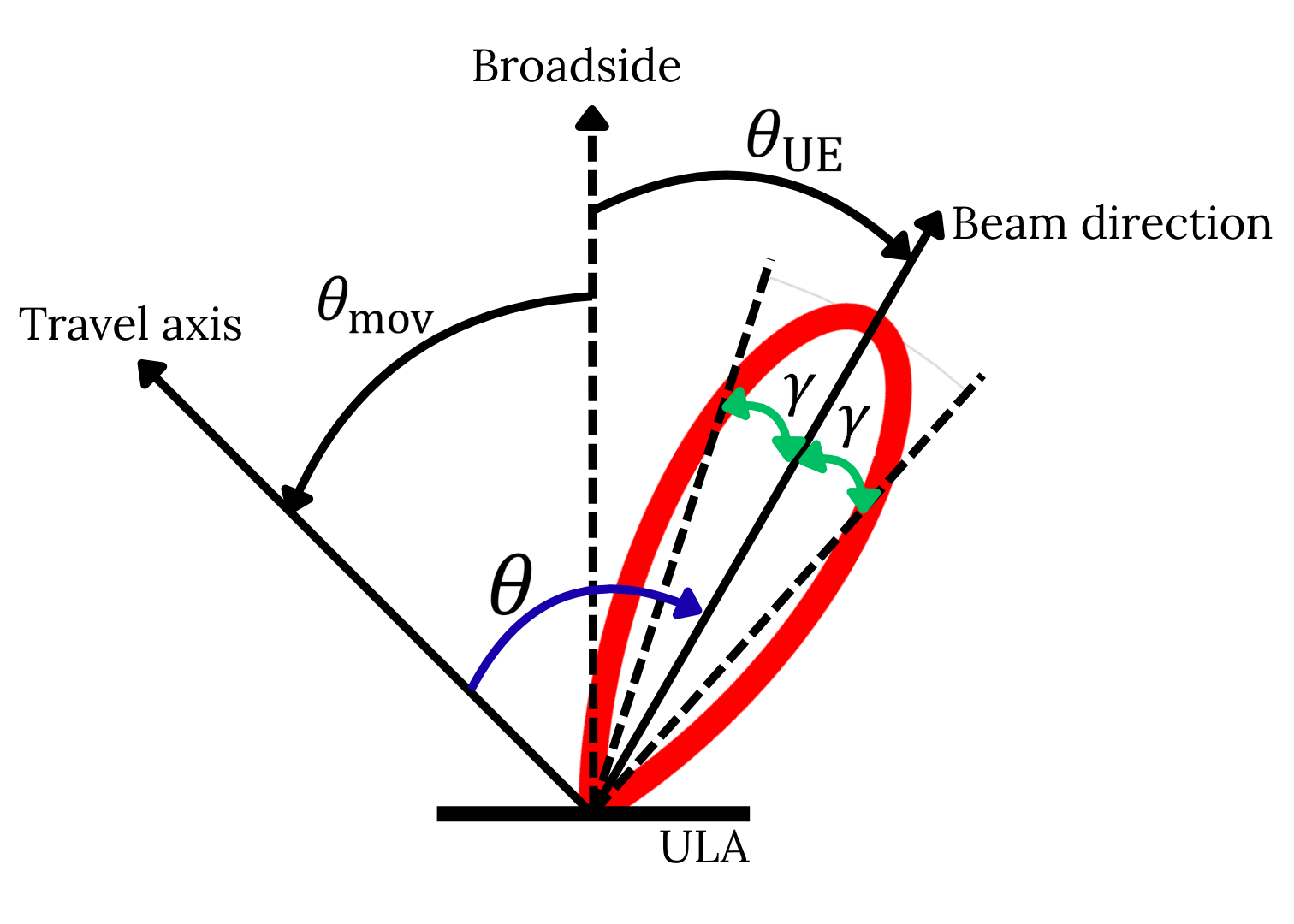}
 \caption{Geometry showing the beam steered at $\theta_{\text{UE}}=\theta+\theta_{\text{mov}}$ with edges at $\theta\pm\gamma$, which determine the Doppler spread.}
 \label{fig:doppler}
\end{figure}

\subsection{Travel-axis outside the main lobe}
We begin by considering the configuration presented in Fig.~\ref{fig:doppler}, where the travel-axis lies outside the UE main lobe, i.e., $\theta\in[-\pi/2-\theta_{\text{mov}},-\gamma]\cup [\gamma, \pi/2-\theta_{\text{mov}}]$. The maximum Doppler spread within the main lobe occurs between its two edges, at angles $\theta+\gamma$ and $\theta-\gamma$ from the travel-axis. Using \eqref{eq:doppler_1}, the maximum Doppler spread is given by
\begin{equation}
\begin{aligned}
\nu_\text{max} \triangleq & \left| \nu_1 -\nu_2 \right| = \frac{\text{v}_\text{UE}}{c}f_c \left| \cos\left( \theta - \gamma\right) - \cos\left( \theta + \gamma\right) \right| \\
%=&\frac{\text{v}_\text{UE}}{c}f_c | \cos\left(\theta \right) \cos\left(\gamma\right) + \sin\left(\theta \right)\sin\left(\gamma \right) \\ &-\left(\cos\left(\theta \right)\cos\left(\gamma \right) - \sin\left(\theta \right)\sin\left(\gamma \right)\right)| \\
=&2\frac{\text{v}_\text{UE}}{c}f_c\left| \sin\left(\theta \right)\sin\left(\gamma \right) \right|.
\end{aligned} 
\label{eq:doppler_2}
\end{equation}
In order to minimize the channel variability in time, we should minimize this maximum Doppler spread. Since \eqref{eq:doppler_2} corresponds to a positive value, we can equivalently consider the squared Doppler spread
\begin{equation}
\nu_\text{max}^2=4\frac{\text{v}_\text{UE}^2}{c^2}f_c^2 \left(\sin^2\left(\theta \right)\sin^2\left(\gamma \right)\right).
\label{eq:doppler_3}
\end{equation}
This squared formulation simplifies differentiation with respect to the pointing angle $\theta$, while maintaining the extrema behavior since squaring is a monotonic increasing function under non-negative input. Taking the derivative with respect to $\theta$, and applying standard trigonometric identities, we get
\begin{equation}
\begin{aligned}
\frac{\partial \nu_\text{max}^2}{\partial \theta}
%&8\frac{\text{v}_\text{UE}^2}{c^2}f_c^2 \sin^2\left(\gamma \right)\sin\left(\theta \right)\cos\left(\theta \right)\\
=&4\frac{\text{v}_\text{UE}^2}{c^2}f_c^2 \sin^2\left(\gamma \right)\sin\left(2\theta \right).
\end{aligned}
\label{eq:doppler_4}
\end{equation}
The stationary points, associated to local extrema, may be found by
finding the roots of \eqref{eq:doppler_4}, which appear at $\theta= n\pi/2, \quad n \in \mathbb{Z}$. By evaluating these values in \eqref{eq:doppler_3}, we can see that at $\theta=\pm\pi/2$ we have maximum spread
 $\nu_\text{max}^2=4\frac{\text{v}_\text{UE}^2}{c^2}f_c^2 \sin^2\left(\gamma \right)$,
where at least one of the values would fall within the considered interval $\theta\in[-\pi/2-\theta_{\text{mov}},-\gamma)\cup (\gamma, \pi/2-\theta_{\text{mov}}]$. Although at $\theta=0$ we have a minimum, $\nu_{\text{max}}=0$, this value lies outside the considered interval. This observation indicates that the optimum $\theta$ should lie within the interval $\theta\in[-\gamma,\gamma]$ considered next.

%The analysis shows that pointing the transmit beam along the travel axis ($\theta=0$) effectively eliminates edge-to-edge Doppler variation, since all directions within the main lobe produce nearly identical velocity projections and therefore nearly identical Doppler shifts. Because the Doppler does not change much across the beam, the effective channel varies very slowly over time and can be regarded as almost constant for longer intervals.  In contrast, when the beam is pointed away from the travel direction, the Doppler term $\cos\left(\theta \right)$ changes significantly across the main lobe, causing a much larger Doppler spread. This leads to faster channel fluctuations and makes the channel less stable, requiring more frequent updates.

\subsection{Travel-axis inside the main  lobe}
%The previous analysis assumed that $\theta>\gamma$, so that both beam edges satisfy $\theta-\gamma > 0$ and $\theta+\gamma > 0$. The Doppler term $\cos\left(\theta \right)$ decreases monotonically across the lobe, which leads to a significant edge-to-edge Doppler variation. 

Let us now consider the case where the travel-axis is contained within the main beam, i.e. $\theta\in(-\gamma,\gamma)$. In this case the worst-case Doppler spread occurs between the travel axis and the farthest main lobe edge, i.e., associated to $ \max(\vert\theta+\gamma\vert,\vert\theta-\gamma\vert)$. 
%\begin{figure}[htbp]
% \centering 
% \includegraphics[width=0.8\linewidth]{Beam_in_with_notations.png}
 %\caption{Geometry showing the beam steered at $\theta$ with edges at $\theta\pm\gamma$, which determine the Doppler spread when $\theta<\gamma$}
 %\label{fig:doppler_in}
%\end{figure}
Assuming without loss of generality that the main lobe edge farthest from the travel-axis corresponds to $\theta+\gamma$,\footnote{The converse case would lead to the same conclusions due to the even symmetry of the cosine function.} the maximum Doppler spread inside the lobe is given by
\begin{equation}  
\begin{aligned}
    \nu_\text{max}&\triangleq\left| \nu\left(\theta+\gamma\right) -\nu\left(0\right) \right|=\frac{\text{v}_\text{UE}}{c}f_c \left| \cos\left( \theta + \gamma\right) - 1 \right|\\
    &=\frac{\text{v}_\text{UE}}{c}f_c \left| \cos\left( \theta \right)\cos\left( \gamma \right) - \sin\left( \theta \right)\sin\left( \gamma \right) -1 \right|.
\end{aligned}
\label{eq:doppler_in_1}
\end{equation}
Considering again the squared maximum Doppler spread
\begin{equation}  
\begin{aligned}\nu_\text{max}^2=\frac{\text{v}_\text{UE}^2}{c^2}f_c^2 \left( \cos\left( \theta \right)\cos\left(\gamma \right) - \sin\left( \theta \right)\sin\left( \gamma \right) -1\right)^2,
\end{aligned}
\label{eq:doppler_in_2}
\end{equation}
differentiating with respect to $\theta$, and reducing the expression through standard trigonometric identities, we get
\begin{equation}
\begin{aligned}
\frac{\partial \nu_\text{max}^2}{\partial \theta}=4\frac{\text{v}_\text{UE}^2}{c^2}f_c^2\left(\sin^2\left(\frac{\theta+\gamma}{2}\right)\sin(\theta+\gamma)\right).
\end{aligned}
\label{eq:doppler_in_3}
\end{equation}
The roots are then at $\theta=n\pi-\gamma$, $\forall n \in \mathbb{Z}$. By evaluating \eqref{eq:doppler_in_2} we can note that $\theta=-\gamma$ corresponds to a minimum since $\nu_{\text{max}}^2=0$, and the next maximum is at $\theta=\pi-\gamma$. However, both extrema lie outside the considered interval, since at $\theta=-\gamma$ we break the assumption that $\vert\theta+\gamma\vert\geq\vert\theta-\gamma \vert$, while for the other extreme we have $\theta=\pi-\gamma\geq\gamma$ since $\gamma\leq \pi/2$. Hence, by the extreme value theorem \cite{boyd} the minimum should lie at the boundary of the interval closest to $\theta=-\gamma$, leading to the optimum value $\theta^*=0$. We have thus proved that the optimum beamforming direction to minimize the Doppler spread, equivalently to maximize the channel coherence, is given by aligning the beam with the travel-axis. This is achieved by selecting $\theta_{\text{UE}}=\theta_{\text{mov}}$ in \eqref{eq:w_UE}.

The previous Doppler analysis is valid for small changes of position, i.e., where the beam maintains visibility of the same scattering objects throughout the trajectory. However, it is also worth noting that, for sufficiently narrow beams pointed far away from the direction of motion, the set of scatterers illuminated by the beam changes rapidly along the UE trajectory, which may create additional channel variability caused by the continual entry and exit of different multipath components. This reinforces the robustness of the proposed travel-axis beamforming, which maintains longer visibility of the scattering objects within the beam.

\section{Numerical Results}

Let us consider the scenario where the UE is positioned at the origin of a coordinate system, as depicted in Fig.~\ref{fig:scenario}, and where the system operates at a carrier frequency of $f_c=10$GHz. The UE is equipped with a four-element ULA oriented perpendicularly to the direction of movement. The BS is positioned $10$m away from the UE in the $\text{x}$-axis and it is equipped with a ULA of $128$ antennas facing the UE, corresponding to a gigantic MIMO configuration.\footnote{Note that the considered 2D scenario only accounts for one of the 2 dimensions of a planar gigantic MIMO array.} The initial UE SNR, defined as $E_s/N_0$, is fixed to $30$dB. Scatterers are present in the scenario, as shown in Fig.~\ref{fig:scenario}. For simplicity, and due to its marginal effect when considering a large number of scatterers in the environment, we assume that scatterers response does not affect the phase, i.e., $\boldsymbol{\alpha}(t)\in \mathbb{R}^{N_{\text{sc}}}$. Their amplitude is then given by the bistatic radar cross section (RCS), which, for the $m-$th scatterer, is approximated as \cite{RCS}
\begin{equation}
\alpha_{m}(t) \approx \text{RCS}_\text{monostatic} \cdot \cos{\left( \frac{\beta_m(t)}{2}\right)},
\label{eq:RCS}
\end{equation}where $\beta_m(t)$ is the bistatic angle for scatterer $m$ with respect to the UE position at time $t$, and the monostatic RCS is assumed constant with $\text{RCS}_\text{monostatic}=0.5$ for all the scatterers.

%\begin{figure}[t]
% \centering 
% \includegraphics[width=1\linewidth]{Scenario.png}
% \caption{Scenario}
% \label{fig:scenario}
%\end{figure}

Two different UE mobility patterns are presented in the evaluation. In the first case, the UE moves perpendicularly from the BS, and in the second one, the UE travels toward the BS along a trajectory forming a $45^\circ$ angle with respect to the BS direction. Both mobility patters are assumed to work under NLOS conditions. For each case, three different beamforming strategies are considered: The first one, omnidirectional transmission at the UE, referred as \textit{Omnidirectional} case. The second is directional beamforming aligned with the UE's direction of motion, denoted as the \textit{Travel-axis} strategy. The third assumes that the UE performs optimal dominant eigenmode transmission by leveraging complete CSI knowledge at the UE, estimated at $t=0$, and is referred to as \textit{Dominant eigenmode} case. Although useful for the comparison, this last case is not practical, since it assumes unsuppressed back‑lobes and requires extensive channel feedback at the UE, conditions that are incompatible with high‑mobility environments.

\subsection{UE moving perpendicular from the BS}

\begin{figure}[htbp]
 \centering
 \includegraphics[width=1\linewidth]{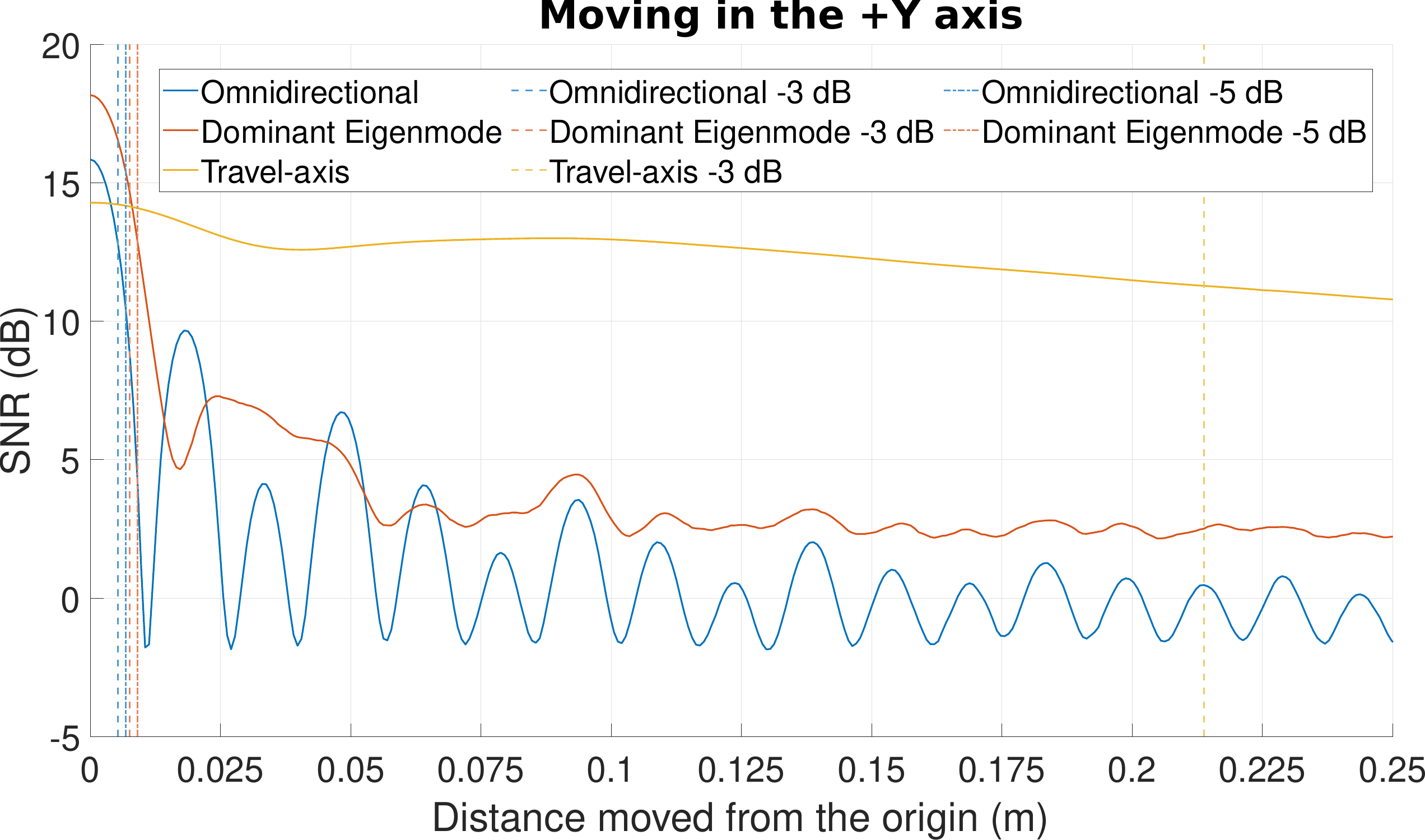}
 \caption{SNR while the UE moves along the Y axis.}
 \label{fig:results_Yaxis}
\end{figure}

For the case where the UE moves along the Y-axis, the results are presented in Fig.~\ref{fig:results_Yaxis}, showing the evolution of the post-combining SNR as a function of the UE movement. When the UE is stationary (step $0$), the \textit{Dominant eigenmode} case provides the highest initial SNR, followed by the \textit{Omnidirectional} case. However, once the UE begins to move, the SNR of both configurations degrades rapidly, whereas the \textit{Travel-axis} strategy maintains a significantly more stable SNR level over a longer distance. Assuming a maximum tolerable SNR degradation of $3$dB, the UE can move up to $5.25$mm in the \textit{Omnidirectional} case, while up to $213.75$mm in the \textit{Travel-axis} case. Additionally, the \textit{Dominant eigenmode} case exhibits a noticeable SNR drop after approximately $7.5$mm of displacement. This behavior further highlights the superior robustness achieved when the beam is aligned with the UE’s direction of movement. Although the initial SNR achieved by the \textit{Travel-axis} strategy is lower than the ones for the \textit{Omnidirectional} and \textit{Dominant eigenmode} configurations, this limitation can be mitigated through techniques such as adaptive gain control (AGC) at the receiver, which compensates for the reduced average SNR while preserving the improved temporal stability provided.

%The SNR achieved by the \textit{Travel-axis} strategy is lower than the ones for the \textit{Omnidirectional} and \textit{Dominant eigenmode} configurations. Nevertheless, this limitation can be effectively mitigated through techniques such as adaptive gain control (AGC) at the receiver, which compensates for the reduced average power while preserving the improved temporal stability provided. 

%Moving $10$m/s ($36$km/h), this corresponds to approximately $1905$ channel characterizations per second in the \textit{Omnidirectional} case, but a more feasible $47$ characterizations per second in the \textit{Pointing} case. For the \textit{Dominant eigenmode} case, it corresponds to $1333$ channel characterizations per second. 

\subsection{UE moving along a $45^\circ$ trajectory}

\begin{figure}[htbp]
 \centering \includegraphics[width=1\linewidth]{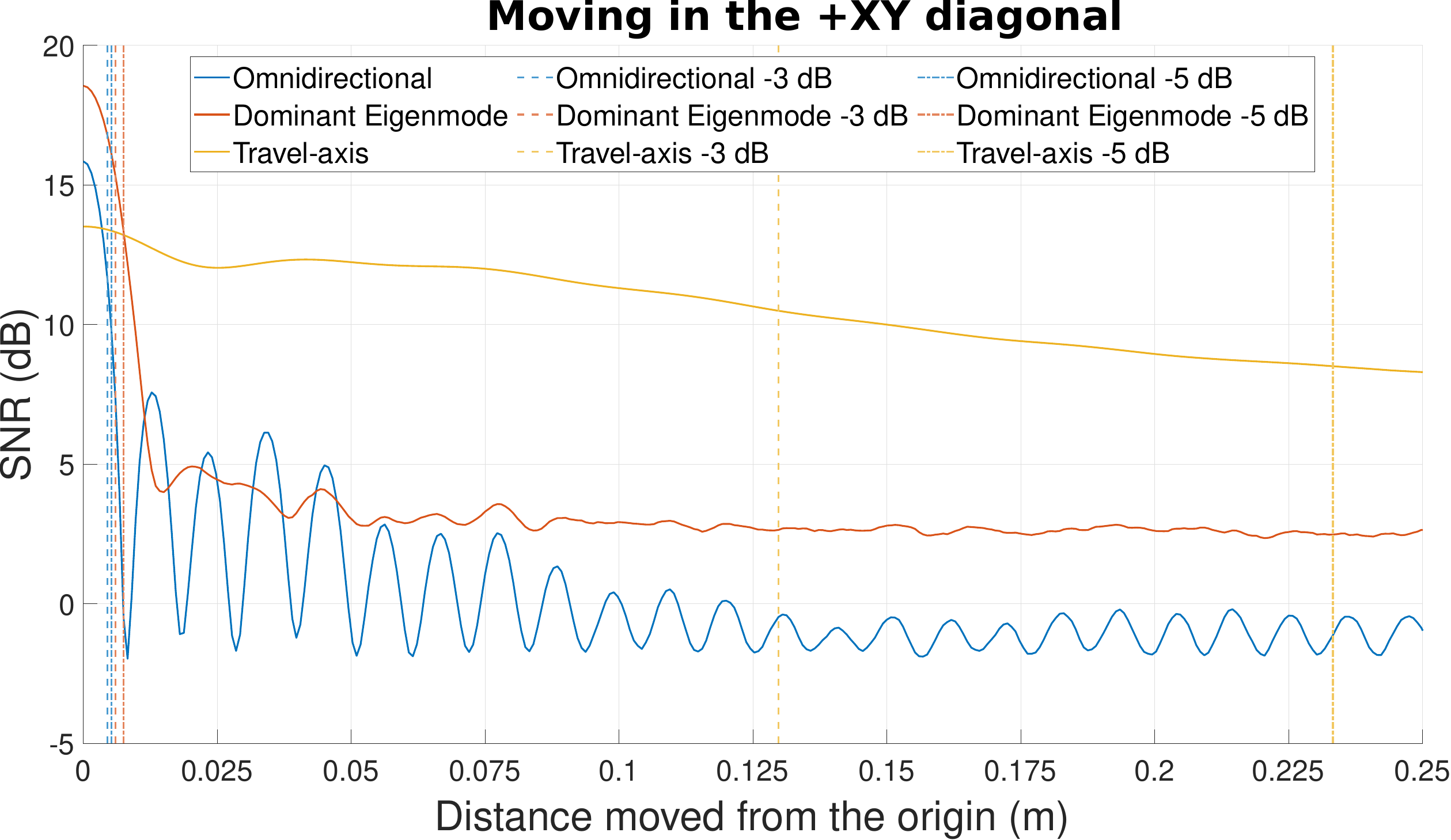}
 \caption{SNR while the UE moves in the +XY diagonal.}
 \label{fig:results_XYaxis}
\end{figure}

A similar behavior is observed when the UE moves towards an angle of $45\degree$ from its origin as shown in Fig.~\ref{fig:results_XYaxis}. In this case, both the \textit{Omnidirectional} and \textit{Dominant eigenmode} configurations achieve a higher SNR than the one achieved by the \textit{Travel-axis} configuration at the initial position, but their performance drops rapidly once the UE begins to move along the trajectory. In contrast, the \textit{Travel-axis} strategy maintains a substantially more stable SNR over a longer displacement, with the $-3$dB loss occurring at a distance of $0.130$m from the origin. 

%Finally, considering that the UE moves towards the BS location, the \textit{Dominant eigenmode} case presents a higher power at the initial position but the $-3$dB loss comes at $0.009$m, which corresponds to $1111$ channel realizations per second. For the \textit{Omnidirectional} case, for this specific case, it corresponds to $1667$ channel realizations per second. Meanwhile, for the \textit{Travel-axis} case, the result decreases to $326$ channel realizations per second. 

%\begin{figure}[htbp]
% \centering \includegraphics[width=1\linewidth]{Xaxis_Paper.eps}
% \caption{SNR while the UE moves towards the BS}
% \label{fig:results_Xaxis}
%\end{figure}

\section{Conclusion and Future Work}

This paper has presented a new UE-side beamforming technique for high-mobility scenarios in NLOS. It demonstrated that, by aligning the beam with the UE trajectory, the temporal evolution of the channel is more stable, reducing the need for frequent channel updates while maintaining reliable communication performance. 

The present study focused exclusively on an UE equipped with a single four‑element array placed perpendicular to the direction of movement. Future work will examine the impact of deploying multiple arrays at the UE and how their relative placement influences channel behavior.

%\begin{thebibliography}{00}
\bibliographystyle{IEEEtran} 
\bibliography{bibliography.bib}
%\printbibliography %Prints bibliography
%\end{thebibliography}
\vspace{12pt}

\end{document}